\documentclass{mem}
\usepackage{natbib}\usepackage{txfonts}\usepackage{balance}
\usepackage{graphicx}
\usepackage[a4paper]{hyperref}
\idline{75}{282}
\begin{document}

\newcommand{\hrd}{HR diagram}
\newcommand{\dtd}{$\Delta$--$t$ diagram}
\newcommand{\msun}{\ensuremath{M_\odot}}
\newcommand{\dov}{\ensuremath{d_\mathrm{ov}}}
\newcommand{\dzz}{\ensuremath{\langle \Delta\nu_0\rangle}}
\newcommand{\tthiz}{\ensuremath{\tilde{t}_\mathrm{HIZ}}}
\newcommand{\ttbcz}{\ensuremath{\tilde{t}_\mathrm{BCZ}}}

\title{
A New Asteroseismic Diagram For Solar-type Stars
}

\subtitle{}

\author{
Anwesh Mazumdar\inst{1,2} 
         }


\institute{
Instituut voor Sterrenkunde, K.\ U.\ Leuven, Celestijnenlaan
200B, 3001 Leuven, Belgium
\and
Observatoire de Paris, LESIA, CNRS UMR 8109, 92195 Meudon, France
}

\authorrunning{Mazumdar}

\titlerunning{New Asteroseismic Diagram}

\abstract{
We propose a new kind of seismic diagram, based on the determination of the
locations of sharp acoustic features inside a star. We show that by combining
the information about the position of the base of the convective envelope or
the \ion{He}{ii} ionisation zone with a measure of the average large
separation, it is possible to constrain the unknown parameters characterising
the physical processes in the stellar interior. We demonstrate the application
of this technique to the analysis of mock data for a CoRoT target star. A
detailed description of this technique is available in \citet{maz05}.

}
\maketitle{}

\section{A new approach --- the \dtd}

The acoustic depth of the \ion{He}{ii} ionisation zone (HIZ) and the base of
the convective envelope (BCZ) can be estimated using the oscillatory signal in
the frequencies that they produce \citep[e.g.,][]{ma01}.  The parameters
\tthiz\ and \ttbcz\ (denoting the acoustic radii of the two layers,
respectively) can be obtained from a least-squares fit of a functional form to
the second differences of the frequencies. For stars more massive than
$1.5\msun$ this method is not reliable due to the proximity of the BCZ and the
HIZ in such stars.

\begin{figure}[t!]
\resizebox{\hsize}{!}{\includegraphics[clip=true]{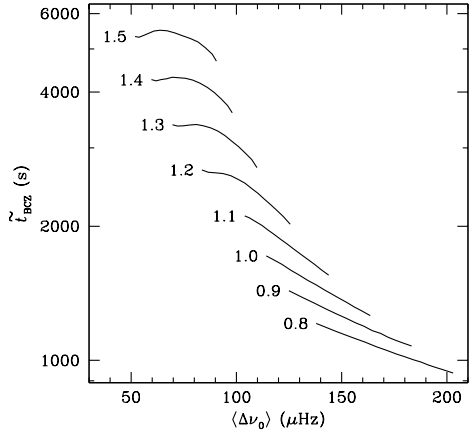}}
\caption{\footnotesize
An example of the \dtd, where the acoustic
radius of the BCZ is plotted as a function of the average large
separation. Each curve on this diagram is an evolutionary track of the
indicated mass in solar units.
\label{fig:delt_mass}
}
\end{figure}

However, the location of the sharp features inside a star (BCZ and HIZ) are not
independent of the general stratification. On the other hand, the mean large
separation is indicative of the gross properties of the star.  In this work we
propose to connect these pieces of information in the form of a diagram (called
the \dtd) to characterise the stellar interior.  We plot the estimated acoustic
radii of the BCZ, \ttbcz, as a function of the mean large separation of the
radial modes, \dzz, for a grid of stellar models (Fig.~\ref{fig:delt_mass}).
Each curve on this diagram is an evolutionary track.  A similar diagram may be
constructed using the acoustic radii \tthiz, instead of \ttbcz.

Clearly, the \dtd\ (Fig.~\ref{fig:delt_mass}) has diagnostic power to determine
the stellar mass and age. Further, the \dtd\ is sensitive to the different stellar
parameters such as mass, chemical composition or convective parameters such as
mixing length and overshoot, which can be exploited to 
search for a suitable model fitting a given set of observed frequencies.
Given an initial trial model, one can test its relative position on the
\dtd\ with respect to the observed data to gain insight into how a
particular parameter needs to be changed from the initial guess in order
to find a better match with the data. This method is
illustrated with an exercise using simulated data for the star HD~49933.

\section{Application of the \dtd}

\begin{figure}[h!]
\resizebox{\hsize}{!}{\includegraphics[clip=true]{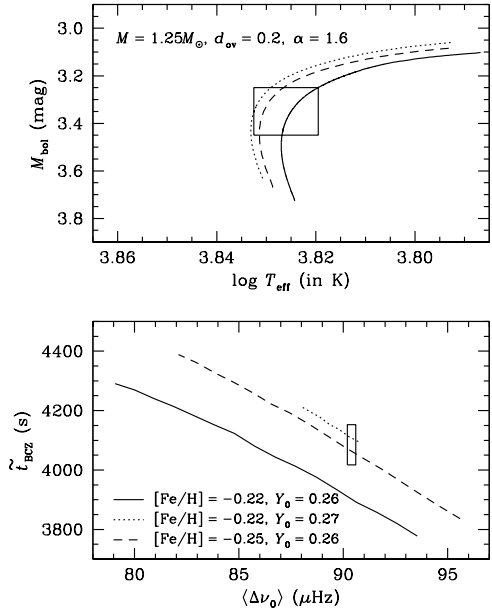}}
\caption{\footnotesize {\it Top panel}: The position of HD~49933 is
shown on the \hrd. Three evolutionary tracks for models with a given
mass, overshoot and mixing length parameter, but different chemical
compositions are illustrated.  {\it Bottom panel}: The \dtd\ for
\ttbcz\ is shown with the three tracks only for the portion where they
lie inside the HR box in the top panel. The derived values of \dzz\ and
\ttbcz\ for HD~49933 are represented by a box.
\label{fig:hd49933_hrd_delt} 
} 
\end{figure}

We have successfully applied the technique of using the \dtd\ to a simulated
dataset of the CoRoT primary target star, HD~49933. The average large
separation was found to be $\dzz = 90.4\pm 0.2~\mu{\mathrm Hz}$. The acoustic
radii were estimated through a fit of the second differences to be $\ttbcz =
4085\pm 68~{\mathrm s}$ and $\tthiz = 4866\pm 49~{\mathrm s}$.  We use this
information to place the star on a \dtd\ constructed from theoretical models.
We start with an initial trial model, characterised by a set of stellar
parameters ($M$, $\alpha$, \dov, ($X_0,Z_0$)), lying inside the error box on
the HR diagram. We place this model on a set of \dtd s, each consisting of a
set of tracks which differ by only one stellar parameter, and compare its
position w.r.t.\ the position of HD~49933. The relative position of the target
star and the trial model on the \dtd\ for each stellar parameter indicates in
which direction that particular parameter needs to be tuned.  An example of the
above procedure is illustrated in Fig.~\ref{fig:hd49933_hrd_delt}, where the
possible values of the chemical composition become constrained. Similar
constraints can be obtained for the other parameters from similar \dtd s.  The
closest model is approached through an iterative procedure of tuning each
parameter, based on the position on the \dtd\ at every step. After several
iterations, we were able to converge to a set of models that satisfied the
constraints on the \dtd\ for both \ttbcz\ and \tthiz.  The task of searching
for the best possible combination of parameters is greatly reduced by the \dtd\
through the additional information about the location of the BCZ and
HIZ.  

\bibliographystyle{aa}

\end{document}